\documentclass[journal,letterpaper]{IEEEtran}

\usepackage{cite}
\usepackage{graphicx}
\usepackage{subfigure}
\usepackage{amsmath}
\usepackage{amssymb}

\begin{document}

%
%

\title{Recent Developments on PIXE Simulation with Geant4}

\author{M. G. Pia,
	G.~Weidenspointner,
        M.~Augelli,
	L.~Quintieri,
        P.~Saracco,
        M.~Sudhakar,	
        and~A.~Zoglauer
\thanks{Manuscript received April 3, 2009.}
\thanks{Maria Grazia Pia and Paolo Saracco are with INFN Sezione di Genova, 
	Via Dodecaneso 33, 16146 Genova, Italy (e-mail:
	MariaGrazia.Pia@ge.infn.it, Paolo.Saracco@ge.infn.it).}
\thanks{Georg Weidenspointner is with the Max-Planck-Institut f\"ur
	extraterrestrische Physik, Postfach 1603, 85740 Garching, Germany, and
	with the MPI Halbleiterlabor, Otto-Hahn-Ring 6, 81739 M\"unchen, Germany
	(e-mail: Georg.Weidenspointner@hll.mpg.de).}
\thanks{Mauro Augelli is with the Centre d'Etudes Spatiales (CNES),
	18 Av. Edouard Belin, 31401 Toulouse, France (e-mail:mauroaugelli@mac.com).}
\thanks{Lina Quintieri is with INFN Laboratori Nazionali di Frascati,
	Via E. Fermi 40, I-00044 Frascati, Italy 
	(e-mail: Lina.Quintieri@lnf.infn.it).}
\thanks{Manju Sudhakar is with INFN Sezione di Genova, Via
	Dodecaneso 33, 16146 Genova, Italy (e-mail:Manju.Sudhakar@ge.infn.it)
	and Phys. Dept., Univ. of Calicut, India; 
	she is on leave from ISRO, Bangalore, India.}
\thanks{Andreas Zoglauer is with the Space Sciences Laboratory,
	University of California at Berkeley, 7 Gauss Way, Berkeley, CA 94720,
	USA (e-mail: zog@ssl.berkeley.edu).}}
\maketitle
\pagestyle{empty}
\thispagestyle{empty}

\begin{abstract}

Particle induced X-ray emission (PIXE) is an
important physical effect that is not yet adequately modelled in Geant4.  
This paper provides a critical analysis of the problem domain
associated with PIXE simulation and describes a set of software developments 
to improve PIXE simulation with Geant4.
The capabilities of the developed software prototype are illustrated
and applied to a study of the passive shielding of the X-ray detectors
of the German eROSITA telescope on the upcoming Russian
\textit{Spectrum-X-Gamma} space mission.

\end{abstract}

\begin{keywords}
Monte Carlo, Geant4, PIXE, ionization.
\end{keywords}

\section{Introduction}
\label{intro}

\PARstart{T}{he} application of particle induced X-ray emission (PIXE) to non-destructive
trace element analysis of materials has first been proposed by Johansson and
co-workers in 1970 \cite{johansson1970}.
Today, this experimental technique is widely exploited in diverse fields 
\cite{pixebook}.

The physical process of PIXE may also give rise to unwanted instrumental
background X-ray lines,
as is the case for space missions and for some laboratory environments.
It also affects the spatial distribution of the energy deposit associated 
with the passage of charged particles in matter: in this respect, its
effects may become significant in the domain of microdosimetry.

The wide application of this experimental technique has motivated the
development of several dedicated software systems; nevertheless,
despite its large experimental interest, limited functionality for
PIXE simulation is available in general-purpose Monte Carlo codes.

This paper discusses the problem of simulating PIXE in the context of a
general-purpose Monte Carlo system  and describes a set of developments 
to improve its simulation with Geant4 \cite{g4nim,g4tns}.
Finally, it illustrates an application of the developed PIXE simulation
prototype to the optimizationof the passive shield of
the X-ray detectors of the eROSITA \cite{erosita} (extended Roentgen
Survey with an Imaging Telescope Array) telescope on the
\textit{Spectrum-X-Gamma} \cite{spectrum} space mission.

\section{Software for PIXE: an overview}

Software tools are available in support of PIXE experimental
applications as specialized codes or included in general-purpose
simulation systems.

\subsection{Specialized PIXE codes} 

Dedicated PIXE codes are focussed on the application of this technique
to elemental analysis. 
They are concerned with the calculation of experimentally relevant
X-ray yields resulting from the irradiation of a material sample by an
ion beam: primarily transitions concerning the K shell,
and in second instance transitions originating from vacancies in the L shell.

For this purpose various analysis programs have been
developed, which are able to solve the inverse problem of determining
the composition of the sample from an iterative fitting of a PIXE
spectrum; some among them are 
GeoPIXE \cite{geopixe},
GUPIX \cite{gupix1,gupix2,gupix3},
PIXAN \cite{pixan}, 
PixeKLM \cite{pixeklm}, 
Sapix \cite{sapix}, 
WinAxil \cite{winaxil} 
and Wits-HEX \cite{witshex}.
A few codes concern PIXE simulation \cite{loh,viba,izarra} specifically.

These codes share basic physics modelling options, like the adoption of the
ECPSSR (Energy-Loss Coulomb-Repulsion Perturbed Stationary State
Relativistic) \cite{ecpssr} model for the calculation of ionization cross
sections; they handle simple experimental geometries, such as target
materials consisting of layers, and impose limitations on the type of
samples that can be analyzed.

\subsection{PIXE simulation in general-purpose Monte Carlo systems}
\label{sec_transport}

Dedicated PIXE software systems have a limited application scope,
as they lack the capability of dealing with complex
experimental configurations.

Comprehensive modelling capabilities are usually associated with
general-purpose Monte Carlo systems.
However, the simulation of PIXE is not widely covered by
large scale Monte Carlo codes treating hadron interactions, while
the conceptually similar simulation of electron impact ionisation is
implemented in the EGS \cite{egsimpact, Kawrakow2006, Hirayama2006} 
and Penelope \cite{penelope} electron-photon Monte Carlo systems.  
The Geant4 simulation toolkit addresses X-ray
emission induced both by electrons and heavy particles like protons
and $\alpha$ particles.

The physics that needs to be considered for the simulation of PIXE
includes the energy loss and scattering of the incident charged
particle, atomic shell ionization cross sections, and atomic
transition probabilities and energies.
On top of these physics features, consistency should be ensured, when
modelling PIXE, with the particle transport schemes governing the
Monte Carlo simulation. 

Intrinsically, PIXE is a discrete process: X-ray emission occurs as the
result of producing a vacancy in atomic shell occupancy, in
competition with Auger electron emission and Coster-Kronig transitions.
Nevertheless, this discrete process is intertwined with the ionization 
process, which determines the production of the vacancy; this process,
for reasons which are elucidated below, is treated in general-purpose
Monte Carlo codes with mixed condensed and discrete transport schemes.
The transport scheme to which ionization is subject affects
the simulation of PIXE.

The simulation of the energy loss of a charged particle due to
ionization is affected by the infrared divergence of the cross section
for producing secondary electrons.
In the context of general purpose Monte Carlo systems, a discrete
treatment of each individual ionization process becomes inhibitive, since, 
due to
their large number, the required computation time becomes excessive.

Infrared divergence is usually handled in general-purpose Monte Carlo
codes by adopting a condensed-random-walk scheme \cite{berger} 
for particle transport. 
In such a scheme, the particle's energy loss and deflection
are treated as averaged net effect of many discrete interactions along
the step, thereby substituting in the simulation a single continuous
process for the many discrete processes that actually occur. 
In a mixed scheme, like the one adopted by Geant4 \cite{g4nim}, two
different r{\'e}gimes of particle transport are introduced, which are
distinguished through a secondary production threshold setting, i.e. a
threshold for the kinetic energy of the electron that is kicked out of
an atom as a result of ionization (the so-called $\delta$-ray): all
ionizations that would generate $\delta$-rays below the threshold are
treated as a continuous process along the step, while the 
ionizations that produce $\delta$-rays above the threshold are
treated as a discrete process.

While this combined condensed-random-walk and discrete particle
transport scheme is conceptually appealing and appropriate to many
simulation applications, it suffers from drawbacks with respect to the
generation of PIXE.

One drawback is that atomic relaxation occurs only 
in connection with the discrete part of the transport scheme, where
the event of producing a $\delta$-ray can be associated with the
creation of a vacancy in the shell occupancy.
For the same reason, the fluorescence yield depends on the threshold 
for the kinetic energy of the secondary electron.

Another drawback of the current transport scheme is that the cross
section for discrete hadron ionization, i.e. for production of a
$\delta$-ray, is calculated from a model for energy loss that 
is independent of the shell where the ionization occurs.
While theoretical calculations are available to determine the spectrum
of the emitted electron for each sub-shell in the case of electron
impact ionization \cite{eedl} for any element, to the authors' knowledge 
no such facilities are currently available for the ionization induced by
protons and ions.
Experimental data are not adequate either to complement the lack 
of theoretical calculations.

\section{Developments for PIXE simulation with Geant4}
\label{sec_newpixe}

At the present time, 
the Geant4 toolkit does not provide adequate capabilities for the
simulation of PIXE in realistic experimental use cases.

The first development cycle 
\cite{saliceti_tesi,relax_nss2004,relax_mc2005,mantero_phd}
had a limited scope: the implementations
concerned only PIXE induced by $\alpha$ particles and involving K
shell ionization - apart from the implementation based on Gryzinski's 
\cite{Gryzinski1965a,Gryzinski1965b}
theoretical model, which produces physically incorrect results.
Even with the models which calculate K shell ionization cross sections
correctly, inconsistencies arise in the simulation of PIXE related to the
algorithm implemented for determining the production of a K shell vacancy.
The deficiencies exhibited by the software released in Geant4 9.2 
\cite{haifa} did not
contribute to improve PIXE simulation with respect to the previous version.

A set of developments for PIXE simulation is described in the following
sections.
A preliminary overview of their progress was reported in \cite{pixe_nss2008}.

The physics aspects associated with PIXE involve the creation of a vacancy 
in the shell occupancy due to ionization, and the emission of X-rays
from the following atomic deexcitation.
The former requires the knowledge of ionization cross sections detailed
at the level of individual shells: for this purpose several  
models have been implemented and validated against experimental data.
The latter exploits the existing functionality of Geant4 Atomic Relaxation
package \cite{relax}.

\begin{figure*}
\centering
\includegraphics[width=15cm]{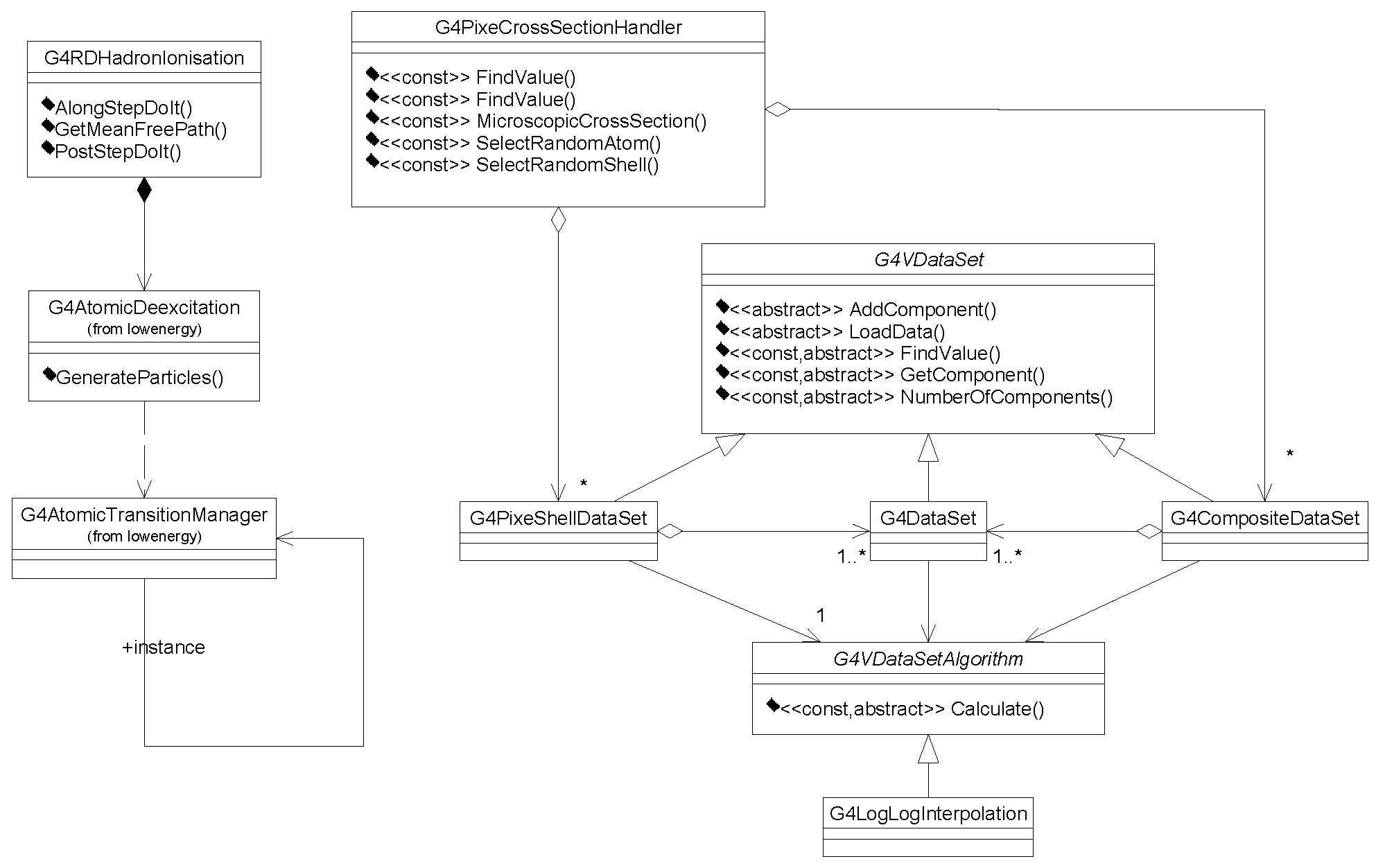}
\caption{The Unified Modelling Language class diagram
of the developments for PIXE simulation described in this paper, 
illustrating the main features of the software.}
\label{UML_fig}
\end{figure*}

The domain decomposition at the basis of PIXE simulation with Geant4
identified three main entities with associated responsibilities: the hadron
ionization process, the creation of a vacancy in the shell occupancy 
resulting from ionisation, the deexcitation of the ionised atom with 
the associated generation of X-rays.
The simulation of PIXE is the result of the collaboration of these entities.
A class diagram in the Unified Modelling Language (UML) \cite{uml}
illustrates the main features of the software design in Fig.~\ref{UML_fig}.

\subsection{Ionisation cross section models}
\label{sec_cross}

The simulation of PIXE concerns a variety of experimental applications,
that require the capability of calculating ionisation cross sections over
an extended energy range: from a few MeV typical of material analysis 
applications to hundreds MeV or GeV range of astrophysical applications.

Various theoretical and empirical models are available in literature
to describe ionisation cross sections for different interacting particles,
as well as compilations of experimental data.
However, there is limited documentation in literature of systematic,
quantitative assessments of the accuracy of the various models.

The current software prototype has adopted the strategy of providing an
extensive collection of ionisation cross section models as a
function of element, atomic (sub-)shell, and incident particle kinetic
energy.

According to the chosen strategy, the provision of a cross section
model is reduced to the construction of tabulations of its values at
preset energies.
The cross sections associated with the models described in this paper
have been pre-calculated either using existing software
documented in literature, or developing ad hoc code.
The data are stored in files, which make up
a data library for PIXE simulation with Geant4 - but could be
used also by other codes.
The cross section data sets selected by the user are loaded into
memory at runtime; cross section values at a given energy are
calculated by interpolation over the tabulated values whenever
required.

The adopted  data-driven approach for the provision of ionization cross
sections presents various advantages. 
It optimizes performance speed, since the calculation of the
interpolation is faster than the calculation from complex algorithms
implementing theoretical models.
This approach also offers flexibility: chosing a cross section model
simply amounts to reading the corresponding set of data files; adding
a new set of cross sections simply amounts to providing the
corresponding set of data files, which are handled transparently.
Finally, the cross section data are transparent to the user: the files are
accessible to the user and human readable.

A wide choice of cross section models for K, L and M shell ionization
is provided in the prototype software for protons and $\alpha$ particles.
The availability of ionization cross section calculations and experimental
data for outer shells is very limited in literature.
Theoretical cross section models
include Plane Wave Born Approximation (PWBA) and variants of
the ECPSSR model: the original ECPSSR formulation \cite{ecpssr},
ECPSSR with United Atom correction (ECPSSR-UA) \cite{isics_ua}, 
ECPSSR with corrections for the Dirac-Hartree-Slater nature of the K
shell \cite{lapicki2005} (ECPSSR-HS), as well as calculations based on
recent improvements to K shell cross section specific to high energy
\cite{lapicki2008} (ECPSSR-HE).  

The cross sections 
have been tabulated and assembled in a data library; the values at a given 
energy are calculated by interpolation.
The tabulations corresponding to theoretical calculations span the
energy range between 10~keV and 10~GeV; empirical models are tabulated
consistently with their energy range of validity.
The adopted  data-driven approach optimizes performance speed and 
offers flexibility for chosing a cross section model.

ECPSSR tabulations have been produced using the ISICS software
\cite{isics,isics2006}, 2006 version and an extended version 
\cite{isics2008} including recent high energy developments.  
Tabulations of ECPSSR calculations as reported in
\cite{paul_sacher} are also provided.

Empirical cross section models for K shell
ionization include the tabulations for protons documented in
\cite{paul_sacher} and a more recent one \cite{kahoul_k}.
An empirical cross section model for K shell ionization by $\alpha$ 
particles is based on the tabulations in \cite{paul_bolik}.
Empirical models for L shell ionization by protons
have been developed by Miyagawa et al. \cite{miyagawa}, Sow et al. 
\cite{sow} and Orlic et al. \cite{orlic_semiemp}.

The ISICS software allows the calculation of cross sections for heavier
ions as well; 
therefore, the current  PIXE simulation capabilities 
can be easily extended in future development cycles.

\subsection{Generation of a vacancy}
\label{sec_vacancy}

The determination of which atomic (sub-)shell is ionised is
related to its ionisation cross section with respect to the total
cross section for ionising the target atom.
However, as previously discussed, the condensed-random-walk
scheme raises an issue as to estimating the total ionisation cross section at a
given energy of the incident particle.

A different algorithm has been adopted with respect 
to the one implemented in the first development cycle: 
the vacancy in the shell occupancy is determined based on 
the total cross section calculated by summing all the individual shell 
ionisation cross sections.
This algorithm provides a correct distribution of the produced vacancies
as long as  ionisation cross sections can be calculated for all the 
atomic shells involved in the atomic structure of the target element.
Since cross section models are currently available for K, L and M
shells only, at the present status of the software this algorithm
overestimates PIXE for elements whose atomic structure involves outer
shells, because of the implicit underestimation of the total ionization 
cross section.
This approach, however,  provides better control on the simulation results
than the algorithm implemented in the first development cycle.

The production of secondary particles by the atomic
relaxation of an ionized atom  is delegated to the
Atomic Relaxation component.

\section{Software validation}

The availability of a wide variety of cross section models for the
first time in the same computational environment allowed a detailed
comparative assessment of their features against experimental data.

The comparison of cross sections as a function of energy was performed
for each element by means of the $\chi^2$ test.
Contingency tables were built on the basis of the outcome of the
$\chi^2$ test to determine the equivalent, or different behavior of
model categories.
The input to contingency tables derived from the results of the
$\chi^2$ test: they were classified respectively as ``pass'' or
``fail'' according to whether the corresponding p-value was consistent
with a 95\% confidence level.
The contingency tables were analyzed with Fisher exact test
\cite{fisher}. 

\subsection{K shell ionization cross sections}
 
\begin{figure}
\centerline{\includegraphics[width=8.5cm]{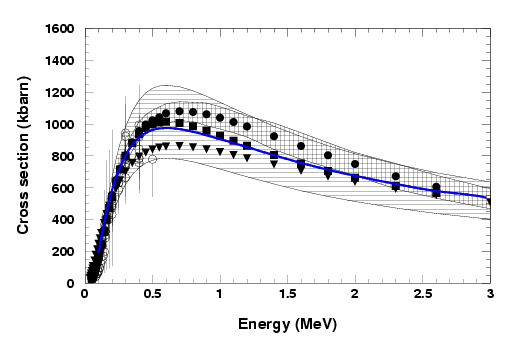}}
\caption{K shell proton ionization cross sections, C: 
ECPSSR model (thick line), with Hartree-Slater (dashed), United Atom 
(dotted) and high energy (thin line) corrections, PWBA (triangles);
Paul \& Sacher \cite{paul_sacher} (filled circles) and 
Kahoul et al. \cite{kahoul_k} (squares) empirical models, 
experimental data from \cite{paul_sacher} (empty circles).
Shaded bands represent  Kahoul et al. (horizontal)
and Paul and Sacher (vertical) uncertainties.
Overlapping curves may not be visible.}
\label{fig_kexp}
\end{figure}

The reference experimental data were extracted from
\cite{paul_sacher}.
An example of how the simulation models reproduce experimental measurements 
is shown in Fig. \ref{fig_kexp}.

\begin{table*}
\begin{center}
\caption{Percentage of test case with compatibility at 
confidence level CL between simulation models and experimental data of K
shell ionization by protons}
\label{tab_kpass}
\begin{tabular}{|c|c|c|c|c|c|c|}
\hline
CL &ECPSSR &ECPSSR-HE	&ECPSSR-HS	&ECPSSR-U  &Paul-Sacher	&Kahoul\\
\hline
\multicolumn{7}{c}{All measurements} \\
\hline
95\%    	&67    &74    &77    &68    &71    &46  \\
99\%		&85    &83    &83    &85    &80    &57  \\
\hline
\multicolumn{7}{c}{Excluding high energy data} \\
\hline
95\%    	&69    &75    &86    &69    &70    &48  \\
99\%		&83    &81    &91    &83    &80    &56  \\
\hline
\end{tabular}
\end{center}
\end{table*}

The fraction of test cases for which the $\chi^2$ test fails to reject the 
null hypothesis at the 95\% and 99\% confidence level are listed in 
Table \ref{tab_kpass}:
all the cross section models implemented in the simulation exhibit
equivalent behaviour regards the compatibility with the 
experimental data, with the exception of the Kahoul 
et al. model.
The contingency table comparing the Kahoul et al.
and ECPSSR-HS models  
confirms that the two models show a statistically significant difference
regards their accuracy (p-value of 0.001).

The contingency tables associated with the other models show that they 
are statistically equivalent regarding their accuracy.
However, when only the lower energy range (below 5-7~MeV, depending on
the atomic number) is considered, a statistically significant
difference at the 95\% confidence level (p-value of 0.034)
is observed between the ECPSSR
model and the ECPSSR-HS one; the latter is more accurate with
respect to experimental data. 

From this analysis one can conclude that the implemented 
K shell ionization cross
section models exhibit a satisfactory
accuracy with respect to experimental measurements.

\subsection{L shell ionization cross sections}
\label{sec_lvalidation}

\begin{figure}
\centering
\subfigure[L$_1$]{\includegraphics[width=8.5cm]{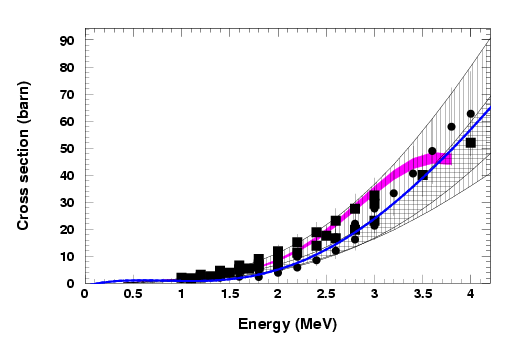}
\label{l1-79}}
%
%
\subfigure[L$_2$]{\includegraphics[width=8.5cm]{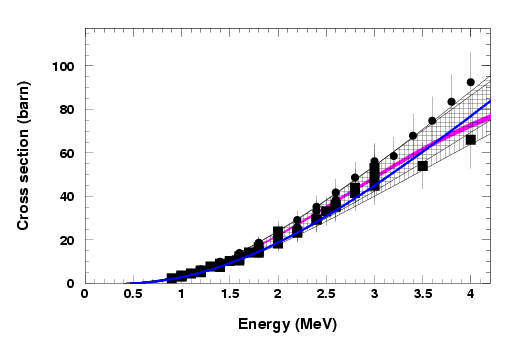}
\label{l2-79}}
%
%
\subfigure[L$_3$]{\includegraphics[width=8.5cm]{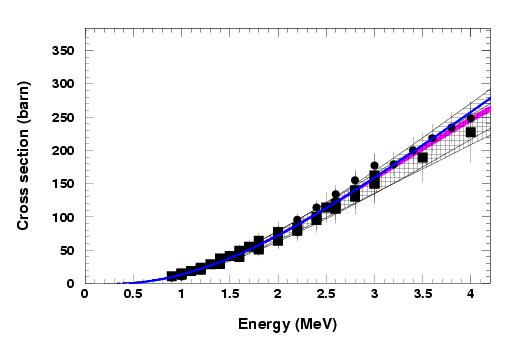}
\label{l3-79}}
\caption{L sub-shell proton ionization cross sections for Z=79:
ECPSSR model (solid line) and ECPSSR with United Atom correction (dashed line);
empirical models by Miyagawa et al. (vertical shaded band), Orlic et al.
(horizontal shaded band) and Sow et al. (solid shaded green band); experimental data 
from \cite{sokhi} (squares) and \cite{orlic_exp} (circles).
Shaded bands represent model uncertainties.
Overlapping curves may not be visible.}
\label{fig_lexp}
\end{figure}

The cross sections for L sub-shell ionization cross sections were compared to
the experimental data collected in two complementary compilations \cite{sokhi},
\cite{orlic_exp}.
An example of how the simulation models reproduce experimental measurements 
is shown in Fig. \ref{fig_lexp}.

The same method was applied as described for the validation of K shell cross
sections.

The ECPSSR model appears to provide a satisfactory representation of
L shell ionisation cross sections with respect to experimental data,
especially with its United Atom variant.

The ECPSSR-UA exhibits the best overall accuracy among the various models;
the Orlic et al. model exhibits the worst accuracy with respect to 
experimental data.  
This semi-empirical
model is the only option implemented in Geant4 9.2 for the calculation of L
shell ionization cross sections.

The accuracy of the various cross section models was studied by means of
contingency tables to evaluate their differences quantitatively.
The categorical analysis was performed between the ECPSSR model with United Atom
correction, i.e. the model showing the best accuracy according to the results of
the $\chi^2$ test, and the other cross section models.
The contingency tables were built based on the results of the $\chi^2$ test at
the 95\% confidence level, summing the ``pass'' and ``fail'' outcome over the
three sub-shells.

The Orlic et al. semi-empirical model is found to be significantly less accurate
than the ECPSSR-UA model: the hypothesis of equivalence of their accuracy with
respect to experimental data is rejected at 99\% confidence level.
The p-values concerning the comparison of the Miyagawa et al. empirical model
are close to the critical region for 95\% confidence, and slightly different for
the three tests performed on the related contingency table.
The Sow et al. empirical model and the ECPSSR model in its original formulation
appears statistically equivalent in accuracy to the ECPSSR model with United
Atom correction.

As a result of this analysis, the ECPSSR model with United Atom approximation
can be recommended for usage in Geant4-based simulation applications as the most
accurate option for L shell ionization cross sections.
The ECPSSR model in its original formulation can be considered a satisfactory
alternative; the Sow et al. empirical model has satisfactory accuracy, but
limited applicability regards the target elements and proton energies it can
handle.

\subsection{Cross section models for high energy PIXE}
\label{sec_hepixe}

PIXE as a technique for elemental analysis is usually performed with proton
beams of a few MeV.
In the recent years, higher energy proton beams of a few tens MeV have been 
effectively exploited too.
hHigh energy protons are a source of PIXE in the space radiation 
environment.

The interest in high energy PIXE has motivated recent theoretical investigations
\cite{lapicki2008} concerning cross section calculations at higher energies.
Despite the emerging interest of high energy PIXE, only a limited set of
experimental data is available above the energy range of conventional PIXE
techniques.

The accuracy of the implemented K shell cross section models was evaluated
against two sets of measurements at higher energy
\cite{denker,pineda}, respectively at 66 and 68 MeV.
The experimental measurement with uranium was not included in the comparison,
since it appears affected by some experimental systematics.

The $\chi^2$ test was performed first separately on either experimental data set
to evaluate the possible presence of any systematics in the two test cases, then
on the combined data set.
The p-values from the $\chi^2$ test against these experimental data are listed
in Table \ref{tab_hepixe}.

Over the limited data sample considered in this test, the ECPSSR model 
with the correction in \cite{lapicki2008}this model does not appear
to provide better accuracy than the original ECPSSR formulation; nevertheless
more high energy experimental data would be required to reach a firm 
conclusion.
Also, this analysis should be verified over tabulation deriving 
from a published version of the ISICS code, when it becomes available. 

The $\chi^2$ test over the experimental data at 160 MeV collected in
\cite{paul_sacher} results in p-values less than 0.001 for all the 
ECPSSR model variants.
The rejection of the null hypothesis could be ascribed either to the deficiency
of the theory or to systematic effects affecting the measurements; further data
would be required for a sound assessment.

\begin{table}
\begin{center}
\caption{P-values from the $\chi^2$ test concerning high energy experimental data}
\label{tab_hepixe}
\begin{tabular}{|l|c|c|c|c|}
\hline
Experimental		&ECPSSR &ECPSSR		&ECPSSR		&ECPSSR \\
data			&	&HE   		&HS		&UA\\
\hline
\cite{denker}, 68 MeV	&0.612	&0.069		&0.054		&0.612 \\
\cite{pineda}, 66 MeV	&0.235	&0.060		&$<0.001$ 	&0.235 \\
Combined		&0.351	&0.020		&$<0.001$ 	&0.351 \\
\hline
\end{tabular}
\end{center}
\end{table}

\section{Application of the PIXE prototype software}
\label{sec_application}

The prototype components for PIXE simulation described in the previous sections
were applied to a study of the passive, graded Z shielding of the X-ray 
detectors of the eROSITA telescope  \cite{erosita} on the upcoming Russian 
\textit{Spectrum-X-Gamma} space mission. 

The purpose of the passive shielding is two-fold.
Firstly, the passive shielding prevents abundant low-energy cosmic-ray
particles from reaching the detectors and thus from causing radiation
damage. 
Secondly, the passive graded Z shielding serves to reduce instrumental
background noise to a minimum \cite{ccd_shielding}. 
This background noise consists of both
fluorescence lines and continuum background due to bremsstrahlung
photons and $\delta$ rays from cosmic-ray interactions.

The event timing accuracy of current imaging Si detectors for X-ray
astronomy (photon energy range $\sim 0.1 - 15$~keV) is limited by the
signal integration time of these devices. For such detectors, an active
anti-coincidence system cannot be used for background reduction
because discarding complete readout frames correlated with an
anti-coincidence signal would result in unacceptable dead
time. Detector triggers due to primary cosmic-ray particles can be
discriminated in imaging detectors due to their high energy deposit
and their pixel image pattern. However, interactions of primary cosmic-ray
particles in the detector and surrounding passive material give rise to
secondary X-rays and charged particles. These may in turn lead to
detector triggers within the accepted energy range. Such triggers
contribute to the instrumental background noise because they cannot be
distinguished from valid events due to cosmic X-ray photons that were
focused by the telescope mirror system onto the detector. 

The production of secondary photons and particles by the ubiquitous
cosmic radiation is inevitable, but graded Z shielding permits the
shifting of the energy of secondaries from atomic deexcitation
to low values. 
The probability that an atomic shell vacancy is filled by radiative
transitions (emission of fluorescence X-ray photons) decreases with
decreasing atomic charge number Z; by contrast, the probability for
non-radiative transitions (emission of Auger and Coster-Kronig
electrons) increases. The average energy of secondaries from atomic
deexcitation decreases with decreasing charge number Z.
Therefore, in a graded Z shield cosmic-ray induced fluorescence X rays
produced in an outer, higher Z layer of the shield are absorbed in an
inner, lower Z layer. 
Subsequent atomic deexcitation in this
inner layer gives rise to fluorescence photons and Auger electrons with energies
that are lower than the energies of the deexcitation particles from the outer
layer; in addition, there will be relatively more deexcitation electrons than
X-rays.
If the effective charge number Z of the innermost shield layer is
sufficiently low, ionization results in the generation of mainly Auger
electrons with energies below 1~keV, which can easily be stopped in a 
thin passivation layer on top of the detectors. 
Ionization can also create rare fluorescence X rays of similarly low energy. 

A first set of graded Z shield designs was studied by Monte Carlo simulation,
using the PIXE prototype software together with Geant4 versions 9.1-patch 01.
The detector chip was modelled as a 450~$\mu$m thick square slab of pure Si with
dimensions 5.6~cm $\times$ 3.7~cm.
The sensitive detector, which is integrated into the chip, covers an area of
2.88~cm $\times$ 2.88~cm or 384 $\times$ 384 square pixels of size 
75~$\mu$m $\times$75~$\mu$m.
This detector model was placed inside a box-shaped shield;
figures of the actual design can be found in \cite{meidinger}.
In its simplest form, the passive shield consisted only of a single 3~cm
thick layer of pure Cu; the outer dimensions of the shield were
12.7~cm $\times$ 10.8~cm $\times$ 7.1~cm. 
A second version of the graded Z shield had a 1~mm thick Al layer inside the Cu
layer, and a third version in addition a 1~mm thick B$_4$C layer inside the Al
layer.
The physics configuration in the simulation application involved the 
library-based processes of the Geant4 low energy electromagnetic package for
electrons and photons, along with the improved version of the hadron 
ionisation process and the specific PIXE software described in 
section \ref{sec_newpixe}. Among the ionization cross section models,
the ECPSSR ones were selected for K, L and M shells.

\begin{figure}[!t]
\centering
\subfigure[Cu Shield]
{\includegraphics[width=8.0cm]{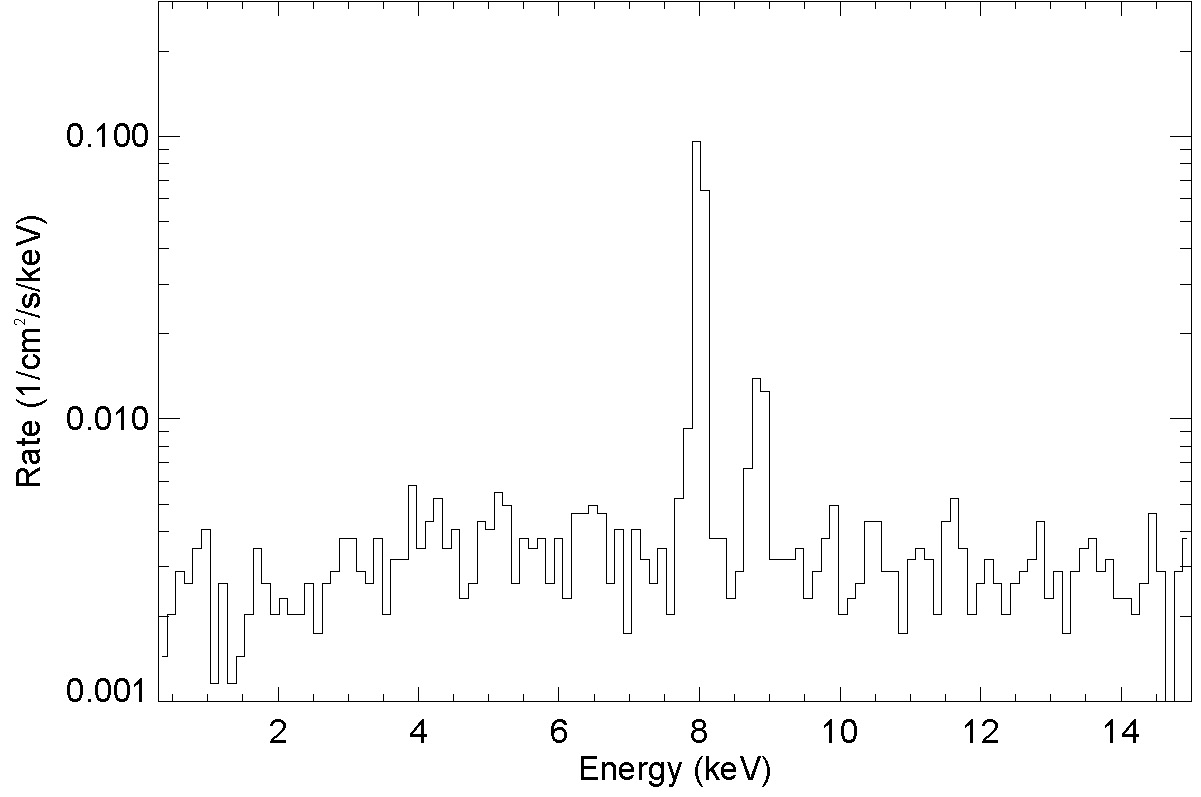}
\label{comp_Cu-shield}}
\vspace*{1ex}
\subfigure[Cu-Al Shield]{\includegraphics[width=8.0cm]{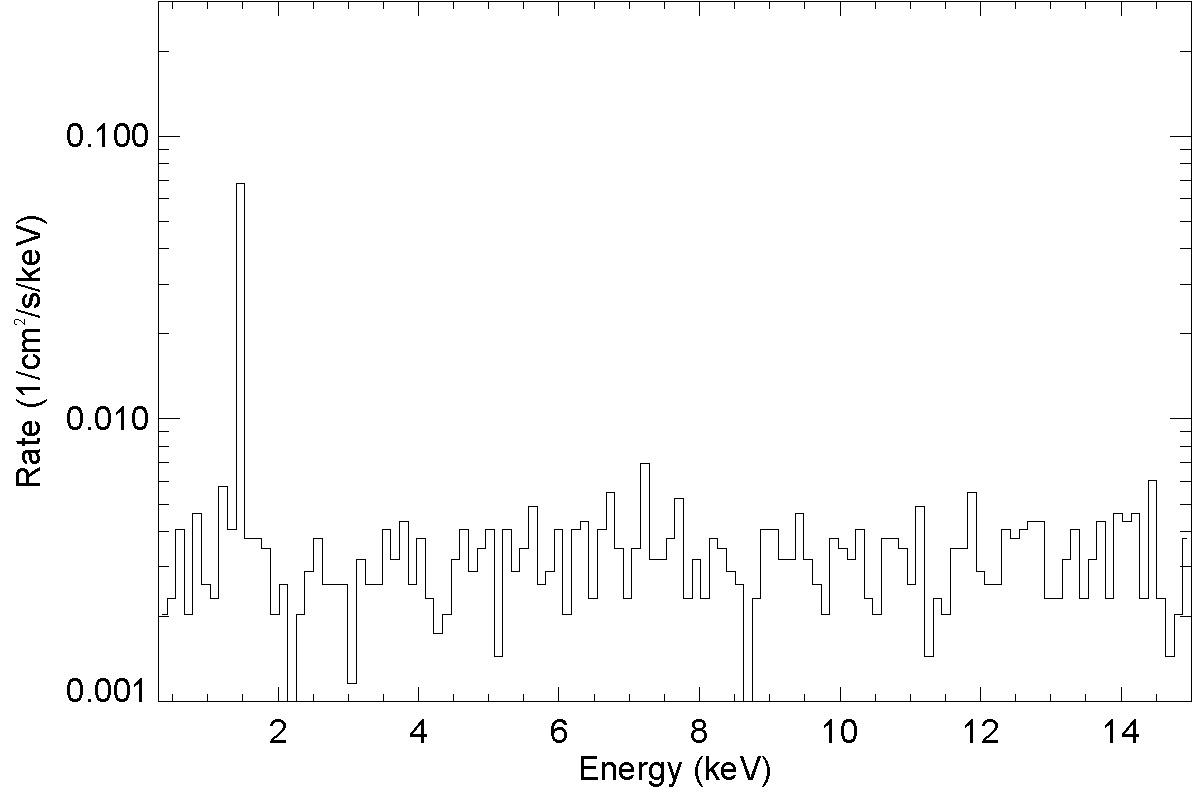}
\label{comp_Cu-Al-shield}}
\vspace*{1ex}
\subfigure[Cu-Al-B$_4$C Shield]
{\includegraphics
[width=8.0cm]{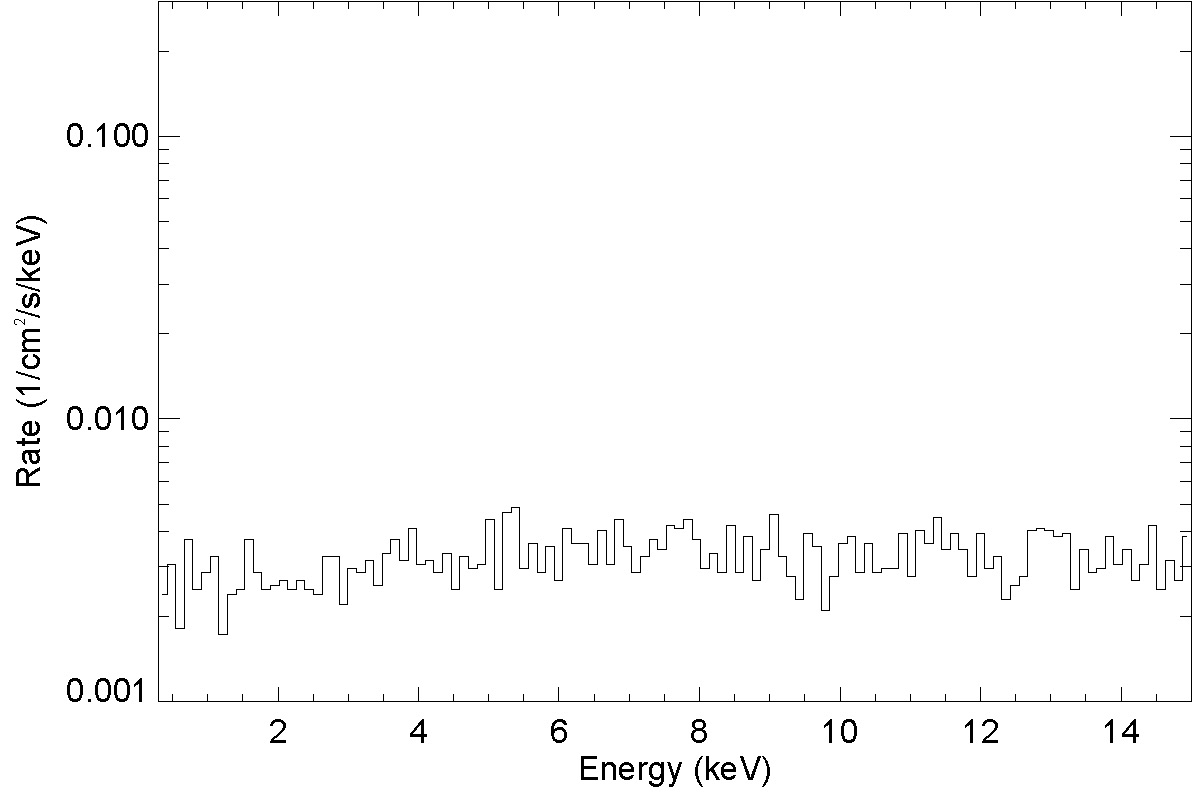}
\label{comp_Cu-Al-B4C-shield}}
\caption{A comparison of the fluorescence background due to ionization
by cosmic-ray protons in an L2 orbit for three different graded Z
shield designs for the eROSITA X-ray detectors.}
\label{comp_shield}
\end{figure}

\textit{Spectrum-X-Gamma} is expected be launched in 2012 into an L2 orbit. 
The background spectra due to cosmic-ray protons were simulated for 
the three different eROSITA graded Z shield designs. 
A model for the detector response, taking into
account Fano statistics (and hence the energy resolution) and detector
noise, was then applied to obtain a simulated data sets. These
simulated data were further processed with specialized data
analysis algorithms \cite{andritschke}.
A comparison of the results is depicted in Fig.~\ref{comp_shield}.
The spectra represent the average background
in a detector pixel. Qualitatively, the PIXE prototype implementation
is working properly: protons
produce the expected fluorescence lines with correct relative
strengths. 
In case of a pure Cu shield,
shown in Fig.~\ref{comp_Cu-shield}, strong Cu K$_\alpha$ and
K$_\beta$ fluorescence lines at about 8.0 and 8.9~keV are present in
the background spectrum. 
In case of a combined Cu-Al shield, depicted in
Fig.~\ref{comp_Cu-Al-shield}, the
Cu fluorescence lines are absorbed in Al, but ionization in Al gives
rise to a clear Al K$_\alpha$ fluorescence line at about 1.5~keV. 
In case of the full graded Z shield configuration, shown in
Fig.~\ref{comp_Cu-Al-B4C-shield}, the B$_4$C layer
absorbs the Al line, but at the same is not a source of significant
fluorescence lines, which is expected due to the low fluorescence
yield of these light elements. 

Except for an excess below 0.3~keV for the case of the Cu-Al-B$_4$C
graded Z shield, which appears because the simulated detector model 
does not yet include a thin passivation layer, 
there is no significant difference in the continuum
background for the three different designs.
The inclusion of a thin layer for the treatment of the low energy background
will be the object of a further detector design optimization.

This application demonstrates that the developed software is
capable of supporting concrete experimental studies.
Nevertheless, the concerns outlined in sections \ref{sec_transport}
and \ref{sec_vacancy}
should be kept in mind: while the present PIXE simulation component
can provide valuable information in terms of relative fluorescence yields
from inner shells, the intrinsic limitations of the mixed transport scheme
in which ionization is modelled and the lack of cross section
calculations for outer shells prevent an analysis 
of the simulation outcome in absolute terms.
.

\section{Conclusion and outlook}
\label{summary}

This paper presents a brief overview of the status, open
issues and recent developments of PIXE simulation with Geant4;
a more extensive report of the underlying concepts, 
developments and results is available in \cite{tnspixe}.

These new developments represent a significant 
step forward regards PIXE simulation with Geant4.
They extend the capabilities of the toolkit
by enabling the generation of PIXE associated with K, L and M shells 
for protons and $\alpha$ particles; for this purpose 
a variety of cross section models are provided.
The adopted data-driven strategy and the software design 
improve the computational performance over
previous Geant4 models.
The validity of the implemented models has been quantitatively
estimated with respect to experimental data.

An extensive ionisation cross section data library has been created as a
by-product of the development process: it can be of interest to the
experimental community for a variety of applications, not necessarily
limited to PIXE simulation with Geant4.

Some issues identified in the course of the development process are
still open: they concern the consistency of PIXE simulation in a mixed
condensed-discrete particle transport scheme.
In parallel, a project \cite{nano5} is in progress to address 
design issues concerning co-working condensed and
discrete transport methods in a general purpose simulation system.

Despite the known limitations related to mixed transport schemes, the
software developments described in this paper provide sufficient
functionality for realistic experimental investigations.


\section*{Acknowledgment}

The authors express their gratitude to A.~Zucchiatti for valuable
discussions on PIXE experimental techniques and to S.~Cipolla for providing 
a prototype version of ISICS including updates
for high energy K shell cross sections.

The authors are grateful to the RSICC staff, in
particular B. L. Kirk and J. B. Manneschmidt, for the support to
assemble a ionization cross section data library resulting 
from the developments described in this paper.


\end{document}